\documentclass{new_tlp}
\pdfoutput=1

\usepackage{mytlpnojournal2}  

\usepackage{myquote}
\usepackage{xspace}
\usepackage{latexsym}
\usepackage{amssymb}   

\usepackage{ifthen}
\newboolean{commentsaon}         
\setboolean{commentsaon}{true}
  \setboolean{commentsaon}{false}  

\newboolean{commentson} 
\setboolean{commentson}{true}

\newcommand{\comment}[1]
{\ifthenelse{\boolean{commentson}\AND\boolean{commentsaon}}
   {{\par\noindent\mbox{}{\small\blue[ *** #1 ]\par}\noindent\par}}{}}

\newcommand{\commenta}[1]
{\ifthenelse{\boolean{commentsaon}}
   {{\par\noindent\mbox{}{\small\color[rgb]{0, .5, 0}[ *** #1 ]\par}\noindent\par}}{}}

\usepackage{pgf}

\usepackage[yyyymmdd]{datetime}

\renewcommand{\today}{2021-08-17}

\usepackage{color}
\newcommand\blue     {\color{blue}}

\usepackage{hyperref}

\newtheorem{theorem}{Theorem}

\newtheorem{lemma}[theorem]{Lemma}

\newtheorem{definition}[theorem]{Definition}

\newcommand*{\seq}[2][n]  {{#2_{1}, \allowbreak \ldots, \allowbreak #2_{#1}}}

\newcommand*{\HU}{{\ensuremath{\cal{H U}}}\xspace}
\newcommand*{\HB}{{\ensuremath{\cal{H B}}}\xspace}
\newcommand*{\TU}{{\ensuremath{\cal{T U}}}\xspace}
\newcommand*{\TB}{{\ensuremath{\cal{T B}}}\xspace}
\newcommand*{\M}{{\ensuremath{\cal M}}\xspace}

\usepackage[mathscr]{euscript}

\newcommand*{\NN}{{\ensuremath{\mathbb{N}}}\xspace}

\newcommand*{\myunderscore}{\mbox{\tt\symbol{95}}}

\newcommand*{\nqueens}{{\sc nqueens}\xspace}

\title[Correctness and completeness]
{On correctness and completeness \\ of an $n$ queens program}

\submitted{. . .}
\revised{\today, Version 6.5.1 }\accepted{. . .}

\author
   [W. Drabent]
   {W{\L}ODZIMIERZ DRABENT%
{\rm\normalsize\small\footnotesize
        {\ifthenelse{\boolean{commentson}\AND\boolean{commentsaon}}
           {\blue\quad  [With private comments]}{}%
        }%
}
   \\ \\
     \normalsize\small
     \begin{tabular}{c}
          Institute of Computer Science,
          Polish Academy of Sciences,\\
          ul. Jana Kazimierza 5,
          01-248 Warszawa, Poland
          \\ 
         and \\
          Department of Computer and Information Science,
          Link\"oping University\\
          S -- 581\,83   Link\"oping, Sweden      
          \\[.5ex]
\email{drabent\,{\it at}\/\,ipipan\,{\it dot}\/\,waw\,{\it dot}\/\,pl}
     \end{tabular}
}

\begin{document}

\maketitle

\begin{abstract}
 Thom Fr\"uhwirth presented a short, elegant and efficient Prolog program for
 the $n$ queens problem.
 However the program may be seen as rather tricky and one may not be convinced
 about its correctness.   
 This paper explains the program in a declarative way, 
 and provides proofs of its correctness and completeness.
The specification and the proofs are declarative, i.e.\ they abstract from any
operational semantics.  The specification is approximate, it is unnecessary 
to describe the program's semantics exactly.  Despite the program works on
non-ground terms, this work employs the standard semantics, based on
logical consequence and Herbrand interpretations.

Another purpose of the paper is to present an example of precise declarative
reasoning about the semantics of a logic program.

\end{abstract}

\begin{keywords}
logic programming, declarative programming, program completeness,
program correctness, specification, non-ground answers
\end{keywords}

\section{Introduction}

 Thom \citeN{Fruehwirth91} presented a short, elegant and efficient Prolog program for
 the $n$ queens problem. %
 However the program may be seen as rather tricky and one may not be convinced
 about its correctness.   The author's description is rather operational.
So it should be useful to explain the program declaratively, and to provide
formal proof that it is correct.

In imperative and functional programming, program correctness implies that
the program produces the ``right'' results.  
In logic programming, which is nondeterministic, the situation is different. 
One also needs the program to be complete, i.e.\ to produce all
the results required by the specification.  (In particular, the empty program
producing no answers is correct whatever the specification is.)

This paper provides proofs of correctness and
completeness of the $n$ queens program; the proofs are declarative,
i.e.\ they abstract from any operational semantics. 

The paper is organized as follows.  After technical preliminaries, 
Section \ref{section.program} presents the $n$ queens program together 
with an informal description of its declarative semantics.  It also 
discusses how to overcome the difficulties with constructing its specification.
The next section presents a formal specification.
Proofs of correctness and completeness of the program 
are subjects of, respectively, 
Sections \ref{sec.correctness.proof} and \ref{sec.completeness.proof}.
Section \ref{sec.comments} discusses the approach.
The last section concludes the paper.

\section{Preliminaries}
\label{sec.preliminaries}

\paragraph{Basics.}
This paper considers definite clause logic programs.
We employ the standard terminology and notation \cite{Apt-Prolog},
and do not repeat here standard definitions and results.
We assume a fixed alphabet of function and predicate symbols.
The Herbrand universe will be denoted by \HU, the Herbrand base by \HB,
and the set of all terms (atoms) by \TU (respectively \TB\/);
$\HB_p$ is the set of ground atoms with the predicate symbol $p$.
By \NN we denote the set of natural numbers.
We sometimes do not distinguish a number $i\in\NN$ from its representation
as a term, $s^i(0)$. 

We use the list notation of Prolog. 
We assume that $[\seq e|e]$ stands for $e$ when $n=0$.
A list (respectively open list) of
length $n\geq0$ is a
term $[\seq e]\in\TU$ ($[\seq e|v]\in\TU$, where $v$ is a
variable); $\seq e$ are the members of
this (open) list.
We generalize the notion of (open) list membership, and say that $e\in\TU$ is
{\bf member} of a term $t\in\TU$ if $t=[\seq[k-1]e,e|e']$, for some terms
$\seq[k-1]e,e'$, where $k>0$.
In such case we also say that $e$ is the $k$-th member of $t$.
Note that this kind of membership is defined by the Prolog built-in predicate
{\tt member}/2.
As in Prolog, each occurrence of \myunderscore\ stands for a distinct variable.

We follow the approach of \citeN{Apt-Prolog} to SLD-resolution.  So we
consider queries instead of goals.  Queries are conjunctions of atoms.
By an {\bf answer} of a program $P$ we mean any query $Q$ such
that $P\models Q$
 ($Q$ is a logical consequence of $P$).
So an answer is a query to which a computed or correct
answer substitution has been applied; \citeN{Apt-Prolog} calls it computed/correct
instance of a query.
(It does not matter
whether correct or computed answer substitutions are considered here, 
due to soundness and completeness of SLD-resolution.) 
\,$\M_P$ stands for the least Herbrand model of a program $P$.
By the {\em relation} defined by a predicate $p$ in $P$ we mean
$
\{\,\vec t\,\in\TU^n \mid P\models p(\vec t\,) \,\},
$
where $n$ is the arity of~$p$.
{}
{}

\paragraph{Specifications.}

In this paper,
the treatment of specifications and reasoning about correctness and
completeness follows that of \cite{Drabent.tocl16};
missing proofs and further explanations can be found there.
For further discussion, examples and references, see also
(\citeNP{Drabent.tplp18}; \citeNP{DBLP:journals/tplp/DrabentM05shorter}).

By a {\bf specification} we mean an Herbrand interpretation $S\subseteq\HB$.
A program $P$ is {\bf correct} w.r.t.\ a specification $S$ when
$\M_P\subseteq S$.  This implies that $S\models Q$ 
($S$ is a model of $Q$)
for any answer $Q$ of $P$.
Note that $S\models Q$ means that each ground instance of each atom of $Q$
is a member of $S$.
A program $P$ is {\bf complete} w.r.t.\ $S$ when  $ S\subseteq\M_P$.
This implies that, for any ground query $Q$, if $S\models Q$ then $Q$
is an answer of $P$.
So $Q$ is an instance of an answer in each SLD-tree for $P$ and any query $Q_0$
more general than $Q$.%
\footnote{%
 As Prolog implements SLD-resolution except for the occur-check, the
 following holds 
 in practice.  If a program $P$, complete w.r.t.\ $S$, is executed as a Prolog
 program with a query $Q_0$, and the computation terminates,
 then all the answers for $Q_0$ described by $S$ are computed.
 More precisely, if $S\models Q$ for a ground instance $Q$ of $Q_0$, 
 then $Q$ is an instance of some answer produced by the computation.
 Similarly, assume that $P$ is correct w.r.t.\ $S$ and is executed with some
 query.  Assume also that 
the occur-check is not needed in this computation
($P$ with the query is occur-check free \cite{Apt-Prolog}).
  Then $S\models Q$ for each
 obtained answer $Q$.  These two facts hold for any selection rule.
}

Dealing with the $n$ queens program we face a usual phenomenon:
Often it is inconvenient (and unnecessary) to specify $\M_P$ exactly, i.e.\ to
provide a 
specification $S$ for which the program is both correct and complete, 
$S=\M_P$.
It is useful to use instead an {\bf approximate specification}, which is a pair
$( S_{\it c o m p l}, S_{corr})$
 of specifications for, respectively,
completeness and correctness.
We say that  
a program $P$ is {\em fully correct} w.r.t.\ 
$( S_{\it c o m p l}, S_{corr})$ when 
$S_{\it c o m p l}\subseteq\M_P\subseteq S_{corr}$.
The approximation is {\em exact} if $S_{\it c o m p l} = S_{corr}$.

 The choice of an approximate specification depends on the properties of
 interest. 
  See for instance
  \cite{drabent2019arxiv.nqueens.forTPLP2021} or \cite{DBLP:journals/tplp/DrabentM05shorter}
  for various specifications for {\sc append}
  describing various properties of the program.

\paragraph{Proving program correctness.}
   An obvious sufficient condition for correctness is provided by 
   Theorem \ref{th.correctness} below. 
   According to \citeN{DBLP:journals/tcs/Deransart93}, the condition is
   due to \citeN{Clark79}.
\begin{theorem}
\label{th.correctness}
    For a program $P$ and a specification $S$,
    if  $S\models P$  then  $P$ is correct w.r.t.\ $S$.
\end{theorem}
\vspace{-1.2\topsep}
\nopagebreak
\begin{proof}
 As $S$ is an Herbrand model of $P$, the least Herbrand model of $P$ is a
 subset of $S$.
\end{proof}

As $S$ is an Herbrand interpretation, \,$S\models P$ means that
     for each ground instance 
    \mbox{$    H\gets \seq B    $}
    ($n\geq0$)
    of a clause of $P$,
     if $\seq B\in S$ then $H\in S$.

\paragraph{Proving program completeness.}
First we introduce some auxiliary notions.
\begin{definition}
  A ground atom $H$ is
  {\bf covered} {by a clause} $C$ w.r.t.\ a specification $S$
  if $H$ is the head of a ground instance  
  $
  H\gets \seq B
  $
  ($n\geq0$) of $C$, such that $\seq B\in S$
  \cite{Shapiro.book}.

  A ground atom $H$ is {\em covered by a program} $P$ w.r.t.\ $S$
  if $H$ is covered w.r.t.\ $S$ by some clause $C\in P$.
\end{definition}
\vspace{-1ex}

\begin{definition}
    A {\bf level mapping} is a function $|\ |\colon\HB\to\NN$.
    A program $P$ is {\bf recurrent} w.r.t.\ a level
    mapping $|\ |$ \cite{DBLP:journals/jlp/Bezem93} 
    when, for each ground instance $H\gets\seq B$ ($n\geq0$) of a clause of $P$
    and each  $i\in\{1,\ldots,n\}$, 
    we have $|H|>|B_i|$.
\end{definition}

The completeness proof presented in this paper will be based on the following
lemma, which is  an immediate corollary  
of \cite[Theorem 5.6 and Proposition 5.4]{Drabent.tocl16}
or of \cite[Theorem 6.1]{Deransart.Maluszynski93}.

\begin{lemma}
\label{lemma.completeness}
Let $P$ be a program, and $S$ a specification.  If 
each atom $A\in S$ is covered by $P$ w.r.t.\ $S$, and
$P$ is recurrent w.r.t.\ some level mapping, then $P$ is complete w.r.t.\ $S$.
\end{lemma}

\paragraph{A note on built-ins.}
The presented approach can be generalized in a rather obvious way to Prolog with
some built-ins.  We focus here on Prolog arithmetic.
A program $P$ using arithmetic predicates (like {\tt is}/2, or {\tt>}/2)
can be understood  \cite{Apt-Prolog} as augmented with an infinite set 
of ground unit clauses
defining the ground instances of arithmetic relations.
Apt denotes the set by $P(Ar)$.
Such clauses are e.g.\,(in the infix form) 
$4\mathop{\tt is}2{+}2$, and 
$2{+}2\mathop{\mbox{\tt<}}7$.
To deal with correctness or completeness of such program, we assume that 
the specification is augmented with $P(Ar)$
(more precisely, that the set of atoms with arithmetic predicates in the
specification is $P(Ar)$).
We also assume that $|B|=0$ for each $B\in P(Ar)$.
Now the sufficient conditions for correctness and completeness apply.
(As they are obviously satisfied by $P(Ar)$,
the condition for correctness
needs to be checked only for the clauses from $P$, and that for completeness
only for atoms with non built-in predicate symbols.)

This approach abstracts from run-time errors.  So completeness w.r.t.\ $S$
means that
if $S\models Q$ and $Q$ is a ground instance of a query $Q_0$
then $Q$ is an instance of an answer of a Prolog computation starting with
$Q_0$, unless a run-time error or infinite loop is encountered.

\section{The $n$ queens program}
\label{section.program}

This section presents the $n$ queens program of
\citeN{Fruehwirth91}, provides its informal declarative description,
and discusses how to construct its specification.
Possible inaccuracies due to informal approach will be corrected in the next
sections, dealing with a formal specification and proofs.

The problem is to place $n$ queens on an $n\times n$ chessboard so that there
is exactly one queen on each row and each column, and at most one queen on
each diagonal. 
 The main idea of the program is to describe 
the position of the queens by a
data structure in which it is impossible that two queens are placed on the
same row, column or a diagonal.
In this way the constraints of the
problem are treated implicitly and efficiently.

This paper considers
the version of the program which represents natural numbers as terms in a
standard way.
Another version employs Prolog arithmetic.
The specifications and proofs of Sections
\ref{sec.spec} -- \ref{sec.completeness.proof} can be, in a rather obvious
way, transformed to ones dealing with the latter version,
following {\em A\,note\,on\,built-ins}\/ from the previous section.

Here is the main part of the program
(with abbreviated predicate names and with the original comment);
it will be named \nqueens.%

\vspace{\abovedisplayskip}
\noindent
\mbox{}\hfill%
\begin{minipage}[t]{.85\textwidth}
{\small %
\begin{verbatim}
    pqs(0,_,_,_).
    pqs(s(I),Cs,Us,[_|Ds]):-
            pqs(I,Cs,[_|Us],Ds),
            pq(s(I),Cs,Us,Ds).

    % pq(Queen,Column,Updiagonal,Downdiagonal)  places a single queen
    pq(I,[I|_],[I|_],[I|_]).
    pq(I,[_|Cs],[_|Us],[_|Ds]):-
            pq(I,Cs,Us,Ds).
\end{verbatim}
}
\end{minipage}%
\hfill
 \begin{minipage}[t]{.035\textwidth}
\raggedleft
\small%
(\refstepcounter{equation}\theequation\label{clause1})
\\ \ \\
       \refstepcounter{equation}%
      (\theequation\label{clause2})%
\\ \ \\ \ \\ \ \\
       \refstepcounter{equation}%
      (\theequation\label{clause3})
\\[1.5ex]
       \refstepcounter{equation}%
        {(\theequation\label{clause4})}
\end{minipage}%
\vspace{\belowdisplayskip}
\\
Solutions to the $n$ queen problem are provided by those answers of \nqueens
that are of the form ${\it p q s}(n,q,t_1,t_2)$,
where $n$ is a number and $q$ a list of length $n$.
A number
 $j\in\{1,\ldots,n\}$
 being the $k$-th member of $q$ means that the queen
    of row $j$ is placed on column $k$.
(The role of $t_1,t_2$ will be explained later.)
\newcommand*{\initialquery}{\ensuremath{Q_{{\rm in},n}}}
    So to obtain the solutions, one can use a query
    $\initialquery = {\it p q s}(n,q_0,\myunderscore,\myunderscore)$,
    where $q_0$ is a list of $n$ distinct variables.

We quote the original description of the program, as it is an example of 
non declarative viewing of logic programs:
\begin{myquote}
  Observing that no two queens can be positioned on the same row, column
  or diagonals, we place only one queen on each row. Hence we can identify
  the queen by its row-number. Now imagine that the chess-board is divided
  into three layers, one that deals with attacks on columns and two for the
  diagonals going up and down respectively. We indicate that a field is
  attacked by a queen by putting the number of the queen there.

  Now we solve the problem by looking at one row at a time, placing one queen
  on the column and the two diagonal-layers. For the next row/queen we use
  the same column layer, to get the new up-diagonals we have to move the 
  layer one field up, for the down-diagonals we move the layer one field down.
\end{myquote}
This does not have much to do with the logic of the program; 
in particular the relations defined by the program are not described.
Instead, actions of the program are described.
Also, the description does not seem to justify why the program is correct.  
  Let us try to treat the program
declaratively, abstracting from the operational semantics.
\paragraph
{Chessboard representation.}
Assume that columns and rows of the $n\times n$ chessboard are numbered 
from 1 to $n$, from left to right and from top to bottom, respectively.
So in an up-diagonal (\rotatebox[origin=c]{-45}{$|$}-diagonal),
all the squares have the same sum of the row number and the column number.
In a down-diagonal (\rotatebox{45}{$|$}-diagonal), the difference of the two
numbers is the same.
Each queen is identified by its row number.

\begin{figure}
\vspace*{-1ex}
\[
  \includegraphics
        [trim = 4.4cm 18.4cm 3.cm 5.3cm, clip, scale=.95] 
  {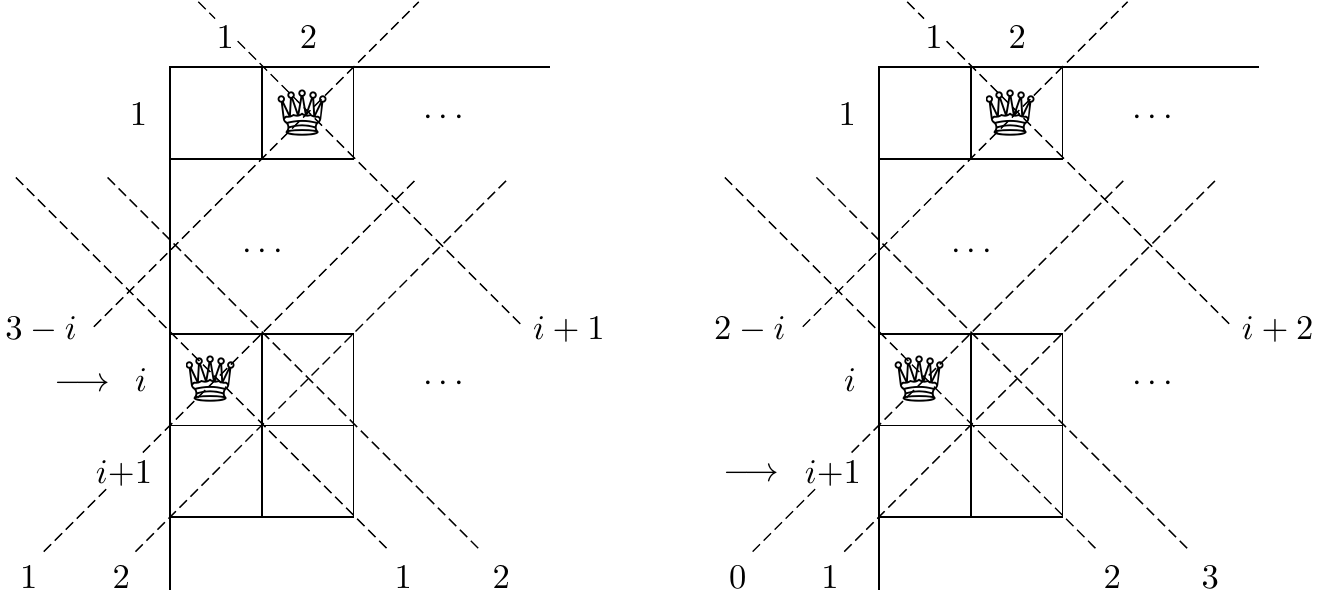}%
\]
  \caption[just a trick]{%
Numbering of rows and columns.  
Numbering of up-diagonals (\rotatebox[origin=c]{-45}{$|$}-diagonals) and
down-diagonals (\rotatebox{45}{$|$}-diagonals)
in the context of row   $i$ (left), and $i+1$ (right), where $i>2$.
}   %
 \figrule
\label{figure.diagonals}
\end{figure}

In contrast to the numbering of rows and columns, 
the numbering of diagonals is not fixed, it is 
specific to the context of the currently considered row;
the diagonal number $m$ includes the $m$-th square of the row, %
for
 $m\in\{1,\ldots,n\}$
(Fig.\,\ref{figure.diagonals}).
So, in the context of row $i$, a queen $j$ (i.e.\ that of row $j$) placed on
a column $k$ is on the up-diagonal of number $k+j-i$,
and on the down-diagonal of number   $k+i-j$.
Consider, for instance, queen $1$ placed on column 2 
(Fig.\,\ref{figure.diagonals}).  Then,
in the context of row $i$, 
it is on the up-diagonal $3-i$, %
and on the down-diagonal $1+i$.    %

Given a set $A\subseteq\{1,\ldots,n\}$ of queens, 
by a {\bf correct placement} of queens $A$ 
we mean placing them on the chessboard so that 
each row, column, up-diagonal and down-diagonal contains at most one queen
from $A$.

When the initial query is \initialquery\ as described above, the program 
represents the position of queens (in the context of a given row $i$) by a list
and two open lists.
In a general case, this representation consists of three terms of the form
$[\seq[l]t|t]$. 
They represent, respectively, the columns, up-diagonals, and down-diagonals,
so that
\begin{equation}
\label{representation}
  \parbox{.88\textwidth}{
  if column (up-diagonal, down-diagonal) $k$, where $k>0$, 
  contains the queen $j$
   \\ then $j$ is the $k$-th member of    
  the term %
  representing the columns  (respectively up-diagonals, down-diagonals).
}
\end{equation}
If a column (or a diagonal) $k$ contains no queen
  then the $k$-th member of the respective term is arbitrary%
\footnote{%
It is a variable when the initial query is \initialquery.
  } %
or does not exist.
Such representation guarantees that at most one queen can be
placed on each column and each diagonal with a positive number.%
\footnote{%
   Diagonals with non-positive numbers are dealt with in contexts in which their
   numbers are positive.
  }

For example,  
a chessboard with two queens $1,i$ 
placed as in Fig.\,\ref{figure.diagonals}
is represented as follows:
The columns are represented by  $[i,1|\ldots]$.
In the context of row $i$,
the up-diagonals are represented by  $[i|\ldots]$
(queen $1$ is not represented here, 
as its up-diagonal has number $3-i\leq0$);
the down-diagonals are represented by $[i,\ldots,1|\ldots]$, where $1$ is the
$(i+1)$-th member of the term
(as the queen 1 is on the down-diagonal number $i+1$). 
In the context of row $i+1$, both queens are not represented in a term 
representing the up-diagonals, and the down-diagonals are represented 
with $[t,i,\ldots,1|\ldots]$, where 1 is the $(i+2)$-th member of the term
(and $t$ is arbitrary).

%
%

  %
  \paragraph{Rationale.}
  Now we informally describe the purpose of the predicates of the program.
  This presentation is preliminary; the aim is to facilitate introducing
  the actual specification.

  The role of ${\it p q}$ is to define  (a relation consisting of)
  tuples $(i, cs, us, d s )\in\TU^4$,
  where
  \begin{equation}
\label{spec.attempt.pq}
\mbox{for some $j>0$, $i$ is the $j$-th member of each $cs,us,d s$.}
\end{equation}
This will be used to assure that a queen $i$ is placed on a column, 
up-diagonal and down-diagonal of the same number.

The role of ${\it p q s}$ is to define
tuples $(i, cs, us, [t|d s] )\in\NN\times\TU^3$
such that
\begin{equation}
\label{spec.attempt.pqs}
\begin{oldtabular}{l@{}}
 $i>0$ and
$cs$,\,$us$,\,$d s$ represent as in (\ref{representation})
\\
a correct placement of queens $1,\ldots,i$
on, respectively, \\
the columns, up-diagonals and down-diagonals, \\
where the diagonals are numbered in the context of row $i$,
\end{oldtabular}
\end{equation}
and additionally all the tuples from $\{0\}\times\TU^3$.
For example, \nqueens has an answer $A$
describing placing two queens on a $4\times4$ chessboard,
\hspace{0pt plus 3pt}%
$A ={\it p q s}(2, cs,us,[\myunderscore|d s] )$, 
\hspace{0pt plus 3pt}%
where
\linebreak[3]
$cs=[1,\myunderscore,2,\myunderscore]$, 
\hspace{0pt plus 3pt}%
$us= [\myunderscore, \myunderscore, 2|\myunderscore]$, 
\hspace{0pt plus 3pt}%
and
$d s= [ \myunderscore, 1, 2|\myunderscore]$.
The argument tuple of $A$ satisfies  (\ref{spec.attempt.pqs}).

  {\footnotesize
  }{}

Now we understand, for instance,
why in clause (\ref{clause2}) the third argument
in the head
${\it p q s( s(I), Cs, Us, [\myunderscore|D s])}$
differs from that in the body atom ${\it p q s(I,Cs,[\myunderscore|Us],D s)}$.
This is because if, according to (\ref{spec.attempt.pqs}), term
 $[\myunderscore|{\it Us}]$ represents
in the context of row $I$
the up-diagonals with positive numbers,
then its tail ${\it Us}$ does this in the context of row $s(I)$.
 Similar reasoning applies to the fourth argument and down-diagonals.

Note that property (\ref{spec.attempt.pqs}) is not closed under substitution.
(E.g.\ answer $A$ above has an instance
  $A'={\it p q s}(2,[1,1,2,\myunderscore],us,d s )$
which places the same queen on two columns, and two queens on a down-diagonal.)
So what we described differs from the relation actually defined by 
${\it p q s}$, and our description needs to be corrected.

\paragraph{Informal specification.}
\hspace{0pt plus .4em}%
Note first that property (\ref{spec.attempt.pq}) is closed under
substitution (due to employing
the generalized notion of member).
Thanks to this our specification for ${\it p q}$ is obvious:
\begin{equation}
\label{spec.pq}
S_{p q} = 
    \{\,
      p q(\,i,\, [\seq[k]c,i|c],\, [\seq[k]u,i|u],\, [\seq[k]d,i|d] \,) 
      \in\HB   %
        {}\mid   k\geq0
      \,\}.
\end{equation}
Note that this specification is exact,
in the sense that $\{(\ref{clause3}),(\ref{clause4})\}$ 
(the fragment of \nqueens defining $p q$) is both correct and complete
w.r.t.\,$S_{p q}$.

The difficulty in constructing a specification for ${\it p q s}$ 
is that the program has
also answers which represent incorrect
 placement of queens.
This cannot be avoided, as any answer of \nqueens like $A$ above
(i.e.\ representing a correct placement of queens with some columns empty)
 has instances which violate the
conditions of the puzzle.
It may seem that we face a contradictory task: the role of our specification
is to describe correct placements, but it has to include some incorrect ones.

The idea to overcome the difficulty is to care only about those atoms
${\it p q s(i, cs, us, d s )}\in\HB$, where $cs$ is a list of distinct members.
It leads to the following informal specification for correctness
for ${\it p q s}$:
\begin{equation}
\label{informal.spec.pqs}  
 \parbox{.55\textwidth}{%
    \hspace*{-1.5em}%
    the set of  those  ${\it p q s(i, cs, us, d s )}\in\HB$ where \\
    $i\in\NN$, \
    $1,\ldots,i$ are members of term  $c s$,
and\\
    if $i>0$ and $cs$ is a list of distinct members then  \\ 
   \mbox{}\quad
    $d s = [t|d s']$ (for some $t,d s'$), and
\\
   \mbox{}\quad condition (\ref{spec.attempt.pqs}) holds for $i,cs,us,d s'$.
}
\end{equation}
It follows that if ${\it cs}$ is a list of length $i$ then 
it is a list of distinct members $1,\ldots,i$, and hence
it is a solution of the $i$ queens problem.
Now our specification for \nqueens is the union of the sets
(\ref{spec.pq}) and (\ref{informal.spec.pqs}).
Note that it serves its purpose, as
correctness w.r.t.\ it implies that the program solves the problem
(each answer for 
the initial query \initialquery\ represents a solution).
Note also that the specification contains atoms which are not answers of
\nqueens,
e.g.\  ${\it p q s}(1,[1,1],[\,],[\,])$.

Such informal specification facilitates understanding of the program and
makes possible informal but precise reasoning
about the program.  
For an example, 
consider a ground instance of clause (\ref{clause2})
\[
\it p q s( s(i), cs, us, [t|d s]) \ \gets \
         p q s(i,cs,[t'|us],d s),\  p q(s(i),cs,us,d s).
\]
Let us denote the body atoms by $B_1$ and $B_2$ respectively.
Assume that they are as described by the specification, i.e.\ 
$B_1\in (\ref{informal.spec.pqs})$,   %
$B_2\in (\ref{spec.pq})$.             %
We show that also the head is as described by the specification, i.e.\ 
is in $ (\ref{informal.spec.pqs})$.
By $B_2\in (\ref{spec.pq})$, $s(i)$ is a member of $cs$.
Assume that $cs$ is a list of distinct members.
So by $B_1\in (\ref{informal.spec.pqs})$ and by (\ref{spec.attempt.pqs}),
 ${\it cs}$, ${\it[t'|{\it us}]}$ and the tail of ${\it d s}$
represent a correct placement of queens $1,\ldots,i$ in the context of row~$i$.
Hence
this placement in the context of row $i+1$ is represented by 
${\it cs, us, d s}$.
By $B_2\in (\ref{spec.pq})$
 we have that, in the same context, $cs,us,d s$ represent
placing the queen $i+1$.
So its column is distinct from those occupied by queens $1,\ldots,i$,
the same holds for its up- and down-diagonals.
Thus ${\it cs, us, d s}$ represent a correct placement of queens $1,\ldots,i+1$.
Hence the head of the clause instance is in the set  (\ref{informal.spec.pqs}).

The reasoning of the last paragraph explains the clause and convinces us
about its correctness. 
Actually it is an informal outline of a central part
of a correctness proof of the program, based on Theorem \ref{th.correctness}.
   In the next section,
   the specification outlined here is made formal and is augmented by a
   specification for completeness.

\section{Approximate specification}
\label{sec.spec}
This section presents a pair of specifications %
for correctness and for completeness of \nqueens,
formalizing the ideas from the previous section.
We often do not distinguish between number $i$ and the queen $i$.

The specification for predicate $p q$ is obvious.  
Both for correctness and for completeness it is
 $S_{p q}$
from (\ref{spec.pq}) in the previous section. 

In order to formulate the specification for ${\it p q s}$, we introduce some
additional notions.  
Assume a queen $j$, i.e.\ that of row $j$,
is in column $k$ of the chessboard described by a list of columns $cs$.
This means that the $k$-th member of $cs$ is $j$.  Then, 
in the context of row $i$, the numbers of the two diagonals containing this
queen are defined as follows.

\begin{definition}
Let a number $j$ be the $k$-th member of a list $cs$.

\quad
The {\bf up-diagonal number} of $j$, w.r.t.\ $i$ in $cs$ is  $k+j-i$.

 \quad
The {\bf down-diagonal number} of $j$, w.r.t.\ $i$ in $cs$ is  $k+i-j$.
\end{definition}

Obviously, queens $j,j'$ are on the same up (respectively down) diagonal iff
for some~$i$
they have the same up (down) diagonal number w.r.t.\ $i$.%
\footnote{%
As an up-diagonal consists of those squares
for which the sum $k+j$ of the column number $k$ and the row number $j$ is
the same. 
Similarly, for a down-diagonal $k-j$ is constant
}
Note that ``for some $i$'' can be replaced by ``for all $i$'', so
we can skip ``w.r.t.\ $i$'' when stating that some queens have distinct up-
(respectively down-) 
diagonal numbers.

\pagebreak[3]

Now we are ready to introduce the core of our specification.

\begin{definition}
\label{def.correct.triple}
A triple of terms $(cs,us,d s)\in\TU^3$  
{\em represents a correct placement} up to row $m$ in the context of row $i$
(briefly: is {\bf correct} up to $m$ w.r.t.\ $i$\/) when $0\leq m\leq i$ and
\[
\qquad\ 
\begin{oldtabular}{l@{}}
  $cs$ is a 
list of distinct members, and each $j\in\{1,\ldots,m\}$ is its member,
\\
  \begin{oldtabular}{@{}l@{\quad}r@{}}
the up (respectively down) diagonal numbers of $1,\ldots,m$ in $cs$ are
distinct, 
&
\refstepcounter{equation}
(\theequation\label{correct.condition.distinct})
\\
          for  each  $j\in\{1,\ldots,m\}$,
          \\
  \quad
  \begin{oldtabular}{l}
      if
      the up (down) diagonal number of $j$ w.r.t.\ $i$ in $cs$ is $l>0$
          \\
      then the $l$-th member of $us$ (respectively $d s$) is $j$.
  \end{oldtabular}
  &
       \refstepcounter{equation}
      (\theequation\label{correct.condition1})
  \end{oldtabular}
\end{oldtabular}
\]
\end{definition}
For example, %
$([1,a,2,b], [c,d,2], [e,1,2])$
is correct up to 2 w.r.t.\ 2, and
the same holds w.r.t.\,3 for
$([1,a,2,b], [d,2],\linebreak[3] [f,e,1,2])$;
both triples represent a correct placement of queens $1,2$ 
on a $4\times 4$ chessboard.
Given that
queens (of rows) $1,\ldots,m$ are placed on the columns as described by $cs$,
condition (\ref{correct.condition1}) assures that  $us,d s$
describe the placement of these queens on the diagonals
with positive numbers.
(An up-diagonal number
in (\ref{correct.condition.distinct}), (\ref{correct.condition1})
may be negative, as  $j\leq i$.)
Obviously, (\ref{correct.condition.distinct}) implies that $cs$ describes a
correct placement of queens $1,\ldots,m$.

\pagebreak[3]

Now the specification for ${\it p q s}$ is
  \[
  \begin{array}[t]{@{}l@{}}
  S_{\it p q s} = 
     \begin{array}[t]{@{}l}
       \big\{\, {\it p q s}(0, cs, us, d s) \mid cs, us, d s \in \HU \,\big\}  
       \ \cup
       \\[1ex]
       \left\{\,  p q s(i, cs, us, [t|d s] ) \in\HB \: \left|\,
       \begin{oldtabular}{l@{\,}}
         $i>0$, \ \ \ %
               $1,\ldots,i$ are members of $cs$,
         \\
         if  $cs$ is a list of distinct members  then \\
          $(cs, us, d s)$ is correct up to $i$ w.r.t.\ $i$. \\
        \end{oldtabular}
        \right\}\right.,
     \end{array}
  \end{array}
\]
and our specification of \nqueens for correctness is
\[
S = S_{\it p q} \cup S_{\it p q s}.
\]

Note that correctness w.r.t.\ $S$ implies the required property of the program.
Take an instance $A$ of query \initialquery, so
 $A = {\it p q s}(n, cs', us', d s')\in\TU$, 
where $n>0$ and
$cs'$ is a list of length $n$.
If $S\models A$  then $cs'$ is a solution of the $n$ queens problem
(as, for each ground instance ${\it p q s}(n, cs, us, u)$ of $A$,
$u$ is of the form $[t|d s]$ and
$1,\ldots,n$ are members of $cs$;
 thus $cs$ is a list of distinct members $1,\ldots,n$, so
  $(cs, us, d s)$ is correct up to $n$ w.r.t.\ $n$,
hence $cs$ represents a solution of the $n$ queens problem).

While specifying completeness, we are interested in ability of the program to
produce all solutions to the problem.  
This leads to the following specification for completeness:
\[
  S_{\it p q s}^0\ =\ 
   \begin{array}[t]{@{}l}
    \left\{\,  p q s(i, cs, us, [t|d s] ) \in\HB \: \left|
     \,
     \begin{oldtabular}{l@{\,}}
       $i>0$, \\
        $(cs, us, d s)$ is correct up to $i$ w.r.t.\ $i$. \\
      \end{oldtabular}
      \right.\right\}.  
   \end{array}
\]

We conclude this section with a property which will be used 
later on.
\pagebreak[3]
\begin{lemma}
\label{lemma.almost.equivalence}
Assume $0<m\leq i$.
Consider two conditions
\[
\begin{tabular}{c@{\qquad}l}
  $(cs,[t|us],d s)$ is correct up to $m$ w.r.t.\ $i$
   &  \refstepcounter{equation}  (\theequation\label{equivalence.part1})
\\
$(cs,us,[t'|d s])$ is correct up to $m$ w.r.t.\ $i+1$
   &  \refstepcounter{equation}  (\theequation\label{equivalence.part2})
\end{tabular}
  \]
For any $t,t'\in\HU$, (\ref{equivalence.part1}) implies
(\ref{equivalence.part2}). 
For any $t'\in\HU$, (\ref{equivalence.part2}) implies 
$\exists\, t\in\HU\:(\ref{equivalence.part1})$. 
\end{lemma}

\vspace{-1\topsep}
\begin{proof}
Assume that $cs$ is a list of distinct members and each $j\in\{1,\ldots,m\}$
is a member of $cs$.  We will consider here the diagonal numbers in $cs$.
Obviously, the up- (down-) diagonal numbers w.r.t.\ $i$ (of $1,\ldots,m$)
are distinct iff the diagonal numbers w.r.t.\ $i+1$ are.

Let $j\in\{1,\ldots,m\}$.
Then $l$ is the down-diagonal number of $j$ w.r.t.\ $i$ 
iff
$l_1=l+1$ is the down-diagonal number of $j$ w.r.t.\ $i+1$.
Note that $l>0$ (as $j\leq i$).
So for down-diagonals,
 conditions (\ref{correct.condition1}) for $i$, $j$, $l$ and $d s$, and    
(\ref{correct.condition1}) for $i+1$, $j$, $l_1$ and $[t'|d s]$ are equivalent.

Number $l$ is the up-diagonal number of $j$ w.r.t.\ $i$ 
iff
$l_2=l-1$ is the up-diagonal number of $j$ w.r.t.\ $i+1$.
So $l_2\geq0$. For $l_2>0$ we, similarly as above, obtain that
for up-diagonals conditions
\vspace{-1.5ex}
\begin{eqnarray}
    &&
    \label{lemma.almost.equivalence.cond.1}
    \mbox{(\ref{correct.condition1}) for $i$, $j$, $l$ and $[t|us]$} 
    \\
    &&
    \label{lemma.almost.equivalence.cond.2}
    \mbox{(\ref{correct.condition1}) for $i+1$, $j$, $l_2$ and $us$}
\end{eqnarray}
are equivalent.
For $l_2=0$, (\ref{lemma.almost.equivalence.cond.1}) vacuously implies
(\ref{lemma.almost.equivalence.cond.2}), and
(\ref{lemma.almost.equivalence.cond.2})
implies that (\ref{lemma.almost.equivalence.cond.1}) holds for some $t$,
namely $t=j$.

This completes the proof of both implications of the lemma.
\end{proof}

\section{Correctness proof}
\label{sec.correctness.proof}

Following Theorem \ref{th.correctness}, 
to prove correctness of program \nqueens w.r.t.\ specification $S$,
one has to show that $S$ is a model of each clause of the program.
In other words to show, for each ground
instance of a clause of the program,
that the head is in $S$ provided the body atoms are in $S$. 
For the unit clauses of \nqueens
\[
\begin{tabular}{l}
${\it p q(I,[I|\myunderscore],[I|\myunderscore],[I|\myunderscore])}$.
\\
${\it p q s}(0,\myunderscore,\myunderscore,\myunderscore)$.
\end{tabular}
\]
it is obvious that each ground instance of the clause is in $S$.
\
Consider clause (\ref{clause4}).
For any its ground instance
\[
{\it p q(i,[t_1|cs],[t_2|us],[t_3|d s])
\gets \it p q(i,cs,us,d s)
}.
\]
it immediately follows from the definition of  $S_{\it p q}$ that
 if the body atom is in $S$ (thus in $S_{\it p q}$) then its head is in 
$S_{\it p q}\subseteq S$.

The nontrivial part of the proof is to show that $S$ is a model of clause 
(\ref{clause2}).  Consider its ground instance
\[
{\it
    p q s(s(i),cs,{\it us},[t|d s]) \gets
            p q s(i,cs,[t_1|{\it us}],d s), \,
            p q(s(i),cs,{\it us},d s).
}
\]
Let $H$ be its head, and $B_1,B_2$ the body atoms.  Assume $B_1,B_2\in S$.
Now (by $B_2\in S$) 
$s(i)$ is the $l$-th member of  $cs, us, d s$ (for some $l>0$).
Note that $l$ is the up (down) diagonal number of $s(i)$ w.r.t.\ $s(i)$
in $cs$.
So condition (\ref{correct.condition1}) holds for  $s(i)$ w.r.t.\  $s(i)$.

Consider first the case of $i=0$.  Then
$(cs, us, d s)$ is correct up to $s(0)$ w.r.t.\ $s(0)$,
provided that $cs$ is a list of distinct members.
Hence $H\in S$.

Consider $i>0$. Note first that $1,\ldots,s(i)$ are members of $cs$
($s(i)$ as explained above, and $1,\ldots,i$ by $B_1\in S$).
Assume that
$cs$ is a list of distinct members.
Then (by $B_1\in S$)
$(cs,[t_1|{\it us}],d s')$ is correct up to $i$ w.r.t.\ $i$, where
$d s '$ is the tail of $d s$.
Hence by Lemma~\ref{lemma.almost.equivalence},
$\beta=(cs,{\it us},d s)$ is correct up to $i$ w.r.t.\ $s(i)$.
As shown above, 
(\ref{correct.condition1}) holds for 
$s(i)$ w.r.t.\ $s(i)$
(where $l$ is both the up- and the down-diagonal number of $s(i)$).
Thus (\ref{correct.condition1}) holds for $1,\ldots,s(i)$ w.r.t.\ $s(i)$.
Hence no up (or down) diagonal number of a $j\in\{1,\ldots,i\}$ is $l$.
As the latter diagonal numbers are distinct (due to $\beta$ being correct up to
$i$), 
(\ref{correct.condition.distinct}) holds for $1,\ldots,s(i)$.

  Hence $\beta$ is correct up to $s(i)$ w.r.t.\ $s(i)$.
  Thus $H\in S$.
  This completes the proof.

\section{Completeness proof}
\label{sec.completeness.proof}

As explained in Section \ref{sec.spec}, we are interested in completeness of
\nqueens w.r.t.\ specification 
$S_{\it p q s}^0$.
However
the sufficient condition of  Lemma \ref{lemma.completeness}
does not hold for this specification.
  Instead let us use 
\[
S^0 = S_{\it p q} \cup S_{\it p q s}^0 \cup 
     \{\, {\it p q s}(0, cs, us, d s) \mid cs, us, d s \in \HU \,\}  
\]
as the specification for completeness.%
\footnote{%
  This is a common phenomenon in mathematics;  an inductive proof of a property
  may be impossible, unless the property is strengthened.
  Actually, the same happened in the case of correctness. We are interested
  in correctness of \nqueens w.r.t.\ $S_{\it p q s}\cup\HB_{\it p q}$.
  However $S_{\it p q s}\cup\HB_{\it p q}$ is not a model of
  the program and Theorem \ref{th.correctness} is not applicable.  Instead
  we used a stronger specification $S = S_{\it p q s} \cup S_{\it p q}$.

  Obviously, correctness (completeness) w.r.t.\ a specification implies
  correctness  (completeness) w.r.t.\ any its superset (subset).
} %

We first show that each atom from specification $S^0$ is covered by program \nqueens.
Each atom
\[
A =  p q(\,i,\, [\seq[k]c,i|c],\, [\seq[k]u,i|u],\, [\seq[k]d,i|d] \,) 
\]
from $S_{\it p q}$ is covered by \nqueens w.r.t.\  $S^0$; for $k=0$ by clause
(\ref{clause3}) as $A$ is its instance;
for $k>0$ by clause (\ref{clause4})
due to its instance 
\newcommand*{\SEQ}[3]
            {{\ensuremath{#1_{#2}, \allowbreak \ldots, \allowbreak #1_{#3}}}}%
$A\gets 
p q(\,i,\, [\SEQ c 2 k,\,i|c],\, [\SEQ u 2 k,i|u],\,\allowbreak
 [\SEQ d 2 k,i|d] \,) 
$ (as its body atom is in $S^0$).
Also, each atom $ {\it p q s}(0, cs, us, d s)$ is covered, as it is an
instance of  clause (\ref{clause1}).

The nontrivial part of the proof is to show that each $A\in S_{\it p q s}^0$
is covered.  Consider such atom, it is of the form
\vspace{-1ex}
\[
A = p q s(s(i), cs, us, [t|d s] ),  %
\]
where $i\geq0$ and $(cs,us,d s)$ is correct up to $s(i)$ w.r.t.\ $s(i)$.
So $cs$ is a list of distinct members, and each $j\in\{1,\ldots,s(i)\}$ is
a member of $cs$.
Let $s(i)$ be the $l$-th member of $cs$.  Thus
$l$ is the up- and down-diagonal number of $s(i)$ w.r.t.\ $s(i)$ in $cs$, and 
(by Definition \ref{def.correct.triple}) $s(i)$ is the $l$-th member of $us$ and
of $d s$.

We show that $A$ is covered by clause (\ref{clause2}) w.r.t.\ $S^0$, due to
its instance 
 \[
 A\gets  B_1, B_2. \qquad
 \mbox{where }
B_1 =  {\it p q s}(i,cs,[t'|us],d s),\ \
B_2 =    p q(s(i),cs,{\it us},d s)
 \]
(and $t'\in\HU$ will be determined later).
We have $B_2\in S^0$ (as $s(i)$ is the $l$-th member of $cs,us$ and $d s$).
If $i=0$ then $B_1\in S^0$, thus $A$ is covered by (\ref{clause2}).%
\footnote{%
    Note that in this case $A$ is covered  w.r.t.\ $S^0$ but not
    w.r.t.\ $S_{\it p q}^0\cup S_{\it p q s}^0$.
    This is why we use $S^0\supset S_{\it p q}^0\cup S_{\it p q s}^0$ as a
    specification. 
}

Assume $i>0$.
As $(cs, us, d s)$ is correct up to $s(i)$ w.r.t.\ $s(i)$, it is correct 
 up to $i$ w.r.t.\ $s(i)$, and 
by Lemma~\ref{lemma.almost.equivalence},
$(cs,[t'|us],d s')$ is correct  up to $i$ w.r.t.\ $i$, for some $t'\in\HU$, 
where $d s'$ is the tail of $d s$.
Hence for such $t'$ we have $B_1\in S_{\it p q s}^0\subseteq S^0$,
thus $A$ is covered by (\ref{clause2}).

This completes the proof that each $A\in S^0$ is covered by \nqueens
w.r.t.\ $S^0$.  
It remains to find
a level mapping under which \nqueens is recurrent.  Consider
the level mapping defined by
\vspace{-1ex}
\[
\begin{array}{l}
  | \, {\it p q s}(i,cs,us,d s) \,| = |i|+|cs|,
\\
  | \, {\it p q}(i,cs,us,d s) \,| = |cs|,
\end{array}
 \qquad \mbox{where} \qquad
    \begin{array}{l}
      |\, [h|t]\, | = 1+|t|,    \\
      |\, s(t)\, | = 1+|t|,    \\
      |f(\seq t)| = 0, 
    \end{array}
\]
for any ground terms $i,cs,us, d s, h,t,\seq t$, and any $n$-ary
 function symbol $f$ distinct from $s$ and from $[\ | \ ]$ ($n\geq0$). 
An easy inspection shows that
under this level mapping \nqueens is recurrent.
Hence by Lemma \ref{lemma.completeness}, the program is complete w.r.t.\ $S^0$.

Additionally, it follows by \cite[Corollary 6.9]{Apt-Prolog} that the
program terminates
for initial queries \initialquery,  %
as each \initialquery\ is bounded w.r.t.\ $|\ |$
 \cite[Definition 6.7]{Apt-Prolog}.
\section{Comments}
\label{sec.comments}

A practical consequence of the correctness, completeness and termination
proven above is as follows. 
Assume that \nqueens with query \initialquery\ 
is executed by Prolog, however with the occur-check.
Then the computation  will terminate, and
the answers will represent all the solutions to the $n$ queens problem.
This holds for arbitrary selection rule.
The occur-check is actually not needed \cite{drabent.tplp2021},
so this also holds for Prolog without the occur-check.

Let us now discuss general applicability of the presented approach.
Note first that finding an appropriate specification was a crucial step to treat
\nqueens within the standard declarative semantics.

Our construction of the approximate specification
exemplifies a general pattern
(\citeANP{Drabent.tocl16} \citeyearNP{Drabent.tocl16,Drabent.tplp18}).
   Some ground atoms may be irrelevant for the program
 properties we are interested in.  
 (In our case they are atoms ${\it p q s(i, cs, us, d s )}\in\HB_{\it p q s}$,
 where $i=0$, or $i\in\NN$, $1,\ldots,i$ are members of $cs$, and
 $cs$ is not a list of distinct members.)
So $\HB$ is divided into a set $S_{\rm i r r}\subseteq\HB_{\it p q s}$
of irrelevant atoms and 
$S_{\rm rel} = \HB \setminus S_{\rm i r r}$, 
the set of relevant
ones.  From the latter a set $S_{\rm c}\subseteq S_{\rm rel}$ of
 ``correct'' ones is chosen,
so that correctness w.r.t.\ specification $S_{\rm i r r}\cup S_{\rm c}$
implies the program properties of interest.
{\sloppy\par}

In many cases, $S_{\rm c}$ describes the ground atoms
which the program should compute, so 
$S_{\rm c}$ is used as a specification for completeness, and 
$S_{\rm i r r}\cup S_{\rm c}$ as one for correctness.

A contribution of our example is dealing with the fact that some
answers of the program represent expected solutions, but
have instances which apparently should be considered
incorrect.  The problem was overcome by including such instances into
$S_{\rm i r r}$.
In this way we do not need to consider non-ground answers, instead it is
sufficient to consider their ground instances from 
$\HB \setminus S_{\rm i r r}$.
This idea should be applicable to other programs posing similar problems.
\section{Conclusions}
 The paper provides an example of precise reasoning about the semantics of a
 logic
 program.  It presents detailed proofs of correctness and completeness of the $n$
 queens program of \citeN{Fruehwirth91}.  
 The program is short, but may be seen as tricky or non-obvious.
 The approach is declarative; the specifications and proofs abstract from any
 operational semantics, the program is treated solely as a set of logical
 formulae. 
 Note that,
 in many cases, approaches based on the operational semantics are proposed 
 for reasoning about declarative properties of logic programs
  \cite{Apt-Prolog,DBLP:conf/tapsoft/BossiC89,DBLP:journals/jlp/PedreschiR99}.
  This seems to introduce unnecessary complications
  (cf.\ \cite[Section 3.2]{DBLP:journals/tplp/DrabentM05shorter}).

The program uses non-ground data,
like open lists with some elements being variables.
Moreover some of its answers (which represent solutions to the $n$ queens
problem) have instances that %
represent incorrect positioning of the queens.
So one may expect that approaches based on the standard semantics and
Herbrand interpretations are inapplicable here.  Actually, this is
not the case. 
We discuss difficulties with constructing a specification, show how to
overcome them, 
and provide a formal specification based on Herbrand interpretations.
Then we prove that the program is correct and complete with respect to the
specification.  
Building the specification is a crucial part of this work.

It may seem that s-semantics \cite{DBLP:journals/tcs/FalaschiLPM89}
is suitable here, as it explicitly deals with non-ground answers.
However the approach employed in this paper seems preferable, as
analogical specification and proofs
employing the s-semantics \cite{drabent2020arxiv.s-semantics} turn out to be
more complicated.

  Our specification is approximate
 (see Section \ref{sec.preliminaries}). 
 Constructing an exact specification of the program would be too
 troublesome, and would result in more complicated correctness and completeness
 proofs.
 This is quite common in logic  programming---one
 often does not need to know the exact semantics of one's program.  
  Some features of the program are of no interest, for instance they may be
  irrelevant to its intended usage.  So we do not need to describe them.

The paper deals with a single program.  In Section \ref{sec.comments}
we discuss the presented ideas as
an instance of a more general approach. 

The detailed proofs presented here may be seen as too impractical due to
numerous details.  
This is however usually the case when proving
program properties.
Experience from imperative and logic programming shows 
 that program correctness really does depend on many 
  details (see e.g.\ the example proofs in
 \cite{AptBO.quicksort2009}
  and in the papers mentioned above).
Maintaining the details by means of some proof assistant is outside of the
scope of this paper. 
On the other hand,
  in the author's opinion proofs like those presented here
  can be informally performed by programmers, %
  possibly in a less detailed way,
  during actual programming.
Two 
fragments of such informal reasoning, with various levels of precision,
are shown in Section \ref{section.program}
(in subsections
{\it Rationale} and {\it Informal specification}).
We expect that formal proof methods, like those discussed here,  can teach
programmers a systematic way
 of reasoning about their programs in practice.

\paragraph{\bf Acknowledgement}
Comments of anonymous referees and of Michael Maher were instrumental in
improving the presentation.

\smallskip\smallskip\noindent
{\it Competing interests}: The author declares none.

\bibliographystyle{acmtrans}
\bibliography{bibpearl,bibmagic,bibs-s,bibshorter}

\end{document}